# Human–Humanoid Collaboration and Ergonomic Risk: An Anthropometric Perspective

He Wen

AI Safety Lab, Bailey College of Engineering & Technology, Indiana State University, Terre Haute, IN 47809, USA

## Abstract

Humanoid robots are increasingly deployed in industrial environments where close physical interaction with human workers is expected. Although these systems are often designed at a human scale, their embodiment is shaped by mechanical, control, and task-oriented constraints rather than biological anatomy. This study examines ergonomic risk in human–humanoid collaboration from an anthropometric perspective using ISO 7250 as a reference framework. Six contemporary humanoid robots are benchmarked based on externally observable geometry to evaluate landmark identifiability, measurement feasibility, and cross-platform patterns of anthropometric deviation. Results show that several ISO-defined landmarks and measurements tied to biological anatomy are consistently inapplicable to humanoid robots, while many geometric and joint-level dimensions remain measurable. Four recurring patterns of anthropometric deviation are observed across platforms: additive, subtractive, exaggerative, and speculative. These findings indicate that ergonomic risk arises not from scale alone, but from how humanoid bodies diverge from human anthropometric assumptions. ISO 7250 remains a useful reference framework, but its direct transfer to humanoid robots is limited, underscoring the need for anthropometry-aware evaluation approaches in collaborative system design.

## 1. Introduction

Humanoid robots are increasingly being introduced into industrial environments where close physical interaction with human workers is unavoidable [1], [2]. Unlike traditional industrial robots that operate in segregated cells, humanoid systems are explicitly designed to share workspaces, tools, and tasks with humans [3], [4]. This shift toward human–humanoid collaboration raises safety and ergonomic challenges that extend beyond conventional concerns in robot control and automation [5], [6], [7]. Ergonomic risk in collaborative systems is fundamentally linked to body geometry, including stature, joint placement, reach envelopes, and mass distribution [8]. In human-centered industrial design, anthropometry provides the basis for evaluating posture, reach efficiency, visibility, and physical strain. However, despite operating on a broadly human scale, current humanoid robots display pronounced diversity in form and substantial anthropometric differences [9]. Even considering height as an example, the height of current humanoid robots varies greatly from 187.00 cm to 5.77 cm, as shown in Figure 1.

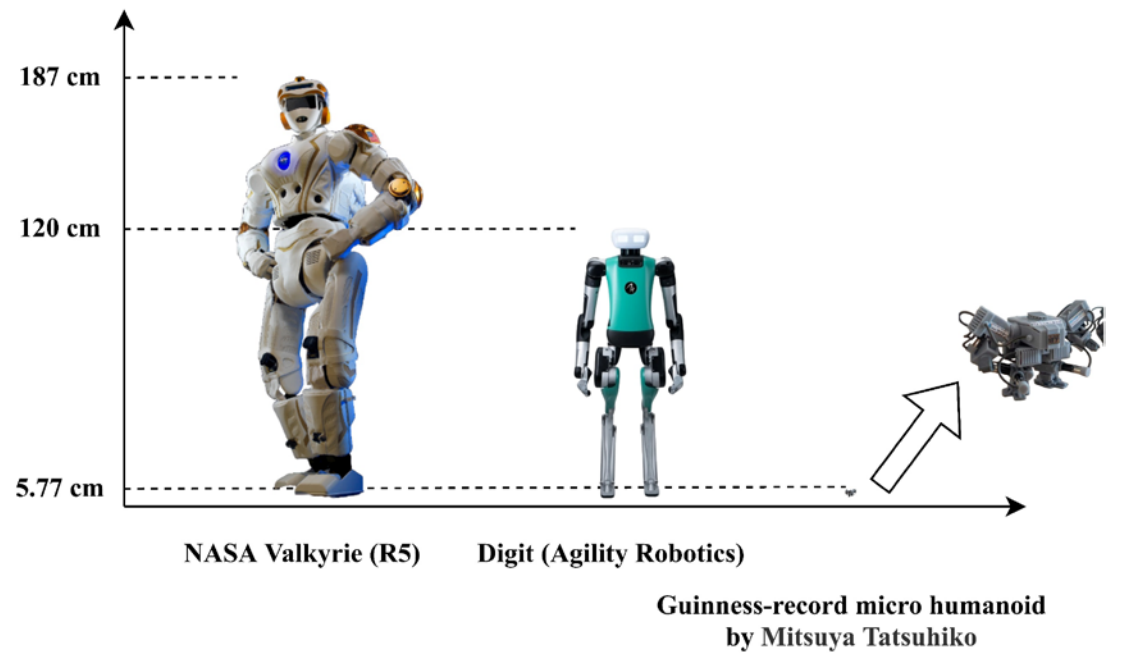

Figure 1: Height comparison of humanoids.

Such differences can substantially affect human–humanoid collaboration and distinguish it from human–human collaboration [10], even when considering anthropometric factors alone. The embodiment of humanoid robots is constrained by actuator packaging, structural design, control requirements, and task-oriented optimization [9]. As a result, humanoid robots may approximate human scale while diverging substantially in joint heights, limb proportions, and external body geometry. Accordingly, ergonomic standards and risk assessment methods implicitly assume biological human body proportions and anatomical landmarks, for example, the *ISO 7250-1:2017 - Basic Human Body Measurements for Technological Design — Part 1: Body Measurement Definitions and Landmarks.*

A further challenge is the limited availability of standardized anthropometric data from robot manufacturers. Unlike human anthropometric databases or industrial equipment specifications, humanoid robots are rarely accompanied by comprehensive dimensional disclosures. This lack of transparency hinders objective evaluation of human–robot physical compatibility and complicates ergonomic risk assessment during system design and deployment. Existing research on human–humanoid collaboration has primarily focused on perception, trust, control strategies, and task allocation [11], [12], [13], while the physical compatibility between human and robot bodies has received comparatively less systematic attention [14], [15]. When ergonomics is discussed, it is often addressed conceptually, without explicit benchmarking of robot anthropometry against human reference frameworks.

Therefore, this study addresses this gap by examining ergonomic risk in human–humanoid collaboration from an anthropometric perspective. Using ISO 7250 as a reference measurement framework, six contemporary humanoid robots are analyzed based on externally observable geometry. Landmark identifiability, measurement feasibility, and patterns of anthropometric deviation are evaluated to clarify where and how human-centered ergonomic assumptions break down when applied to humanoid robots.

## 2. Methodology

This study employs a comparative anthropometric benchmarking methodology to investigate ergonomic risks in human–humanoid collaboration under conditions of limited manufacturer disclosure. As illustrated in Figure 2, the methodology consists of four sequential steps: observe, benchmark, review, and conclude. Each step serves a distinct epistemic function, progressively transforming raw visual evidence into anthropometric safety insights.

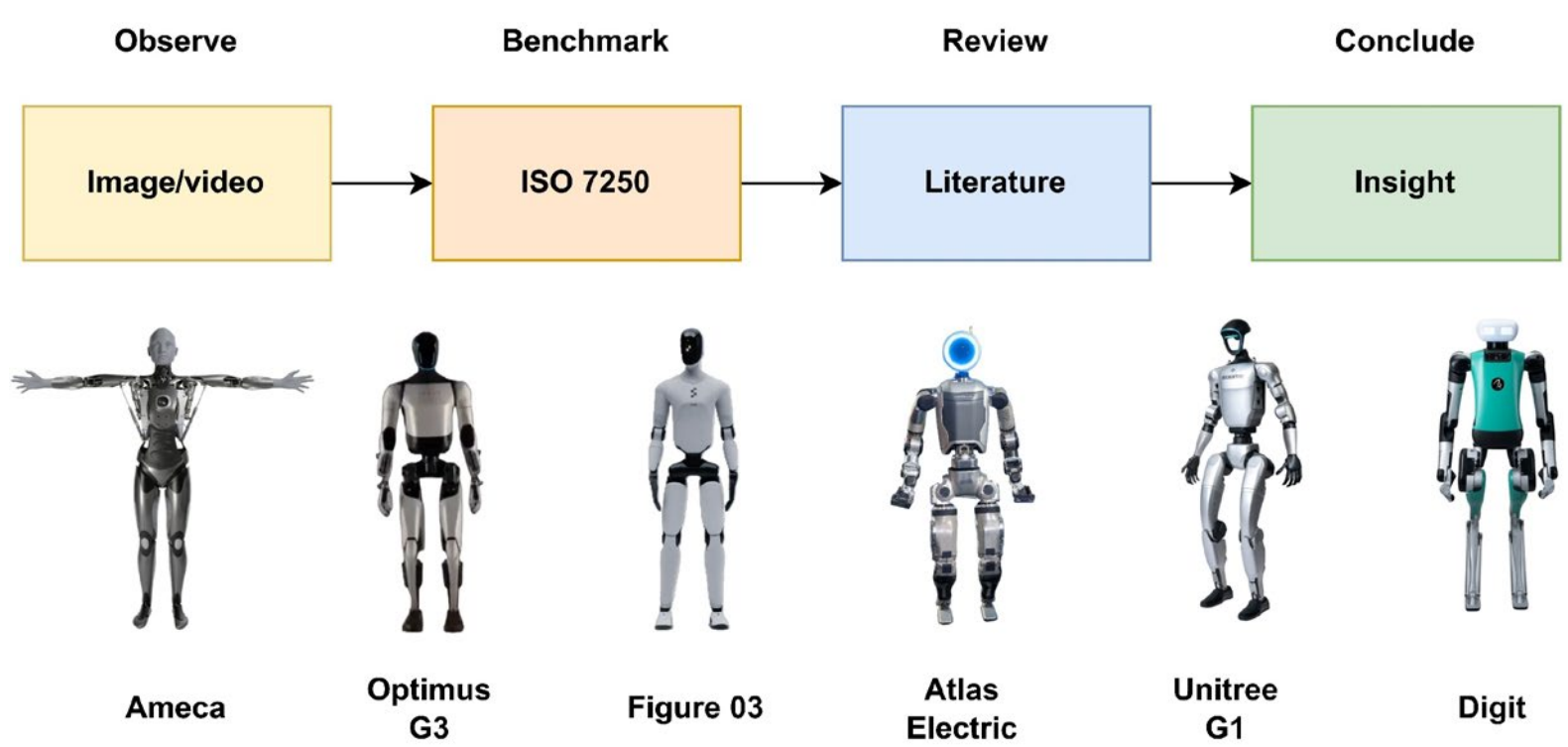

Figure 2: Methodological framework.

Six contemporary humanoid robots (N = 6) were analyzed using ISO 7250 as a reference framework based on publicly available images and videos: Ameca by Engineered Arts, Optimus Gen 3 by Tesla, Figure 03 by Figure AI, Atlas Electric by Boston Dynamics, Unitree G1 by Unitree, Digit by Agility Robotics. Externally observable geometry, including stature, joint placement, and limb segmentation, was used to assess landmark identifiability, proxy-identifiability, and measurement feasibility without relying on internal specifications. Feasible dimensions were benchmarked using normalized and joint-level comparisons to enable cross-platform analysis. The study adopts a structured, reproducible observational benchmarking approach. Observed deviations from human anthropometric assumptions were interpreted using established ergonomic principles, for example, Rapid Upper Limb Assessment (RULA)/Rapid Entire Body Assessment (REBA), and reach/clearance considerations, serving as analytical anchors to infer potential ergonomic risks despite the absence of task-specific posture data.

# 3. Results

## 3.1 Landmark Identifiability

Landmark identifiability was evaluated by determining whether ISO 7250 Clause-5 anthropometric landmarks could be located on humanoid robots using externally observable geometry only, without inferring internal structure or biological equivalence. Each landmark was classified as identifiable, proxy-identifiable, or not identifiable based on visual inspection of publicly available images and videos.

Results show that landmarks associated with rigid body extrema, mechanical joints, or clear geometric transitions (e.g., vertex, acromion, olecranon) were generally identifiable or proxy-identifiable across all robots, enabling consistent joint-level and whole-body measurements. In contrast, landmarks tied to biological skeletal or soft-tissue anatomy, such as iliospinale anterius (ASIS), thelion, tibiale, and tragion (Figure 3), were universally non-identifiable, rendering related measurements infeasible.

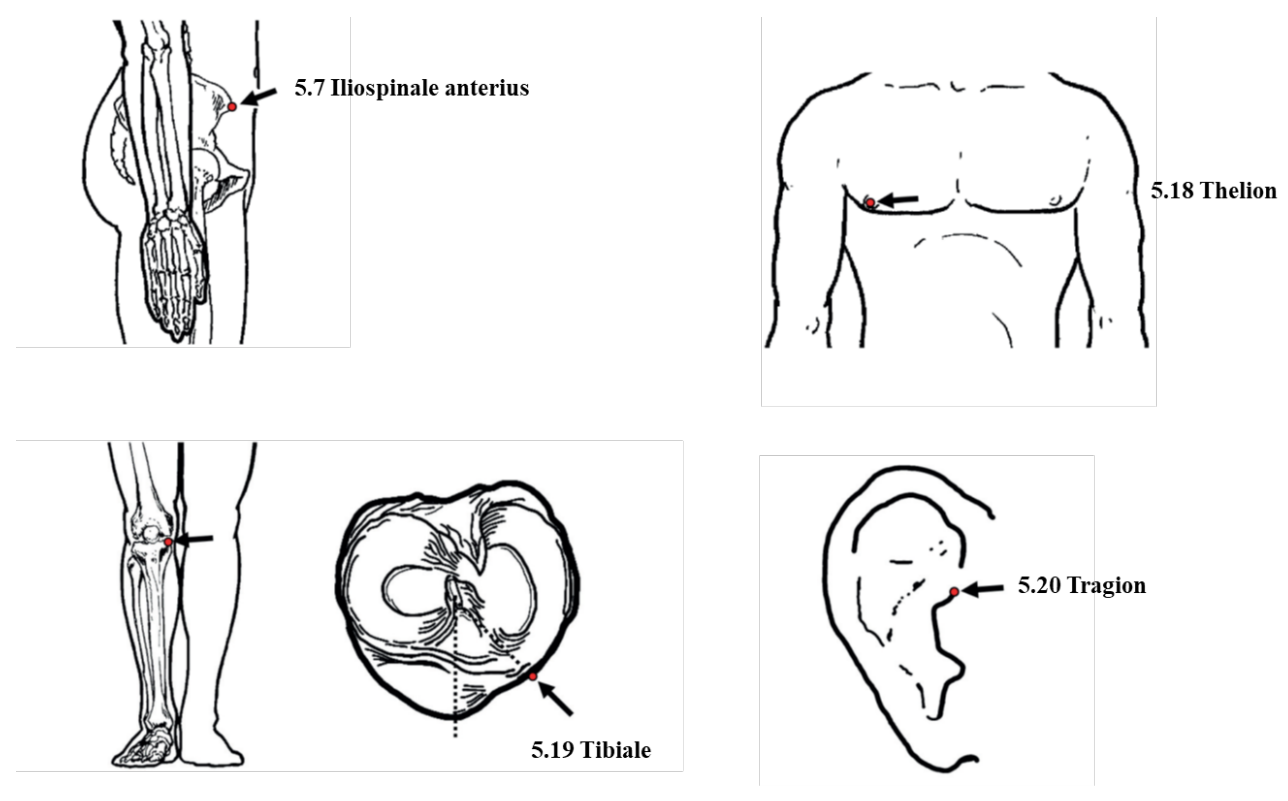

Figure 3: Non-identifiable landmarks.

Facial landmarks were identifiable only on robots with explicitly anthropomorphic facial geometry (e.g., Ameca) and absent on function-oriented designs (Figure 4). These findings demonstrate a structural mismatch between human anatomical landmarks and humanoid robot embodiments and define the practical limits of applying ISO 7250 landmarks under external observation constraints.

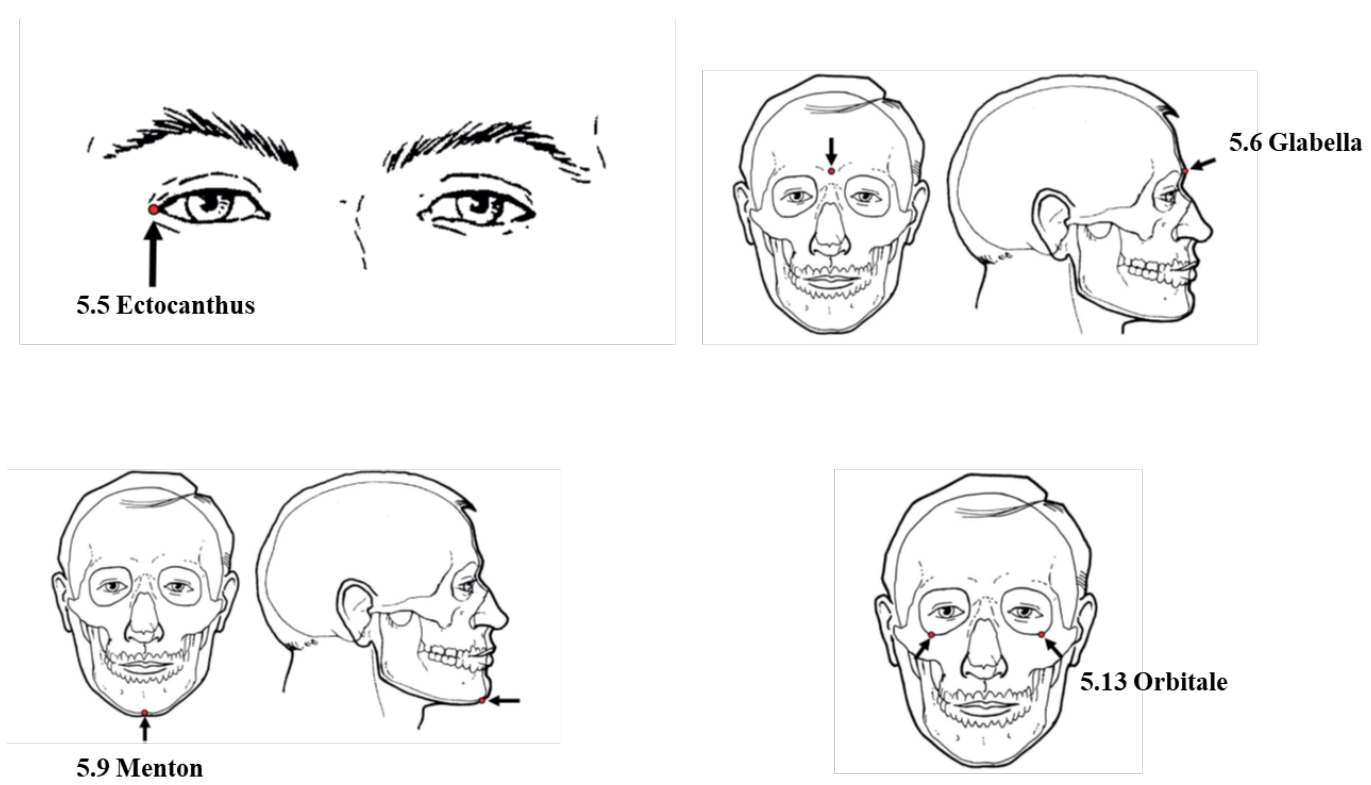

Figure 4: Facial landmarks (most lack, only Ameca has).

**3.2 Measurement Feasibility**

Measurement feasibility was assessed by determining whether ISO 7250 body dimensions could be derived for humanoid robots using identifiable or proxy-identifiable landmarks based solely on externally observable geometry. A measurement was considered feasible only when all required reference points could be located without relying on inferred anatomy or internal structure.

Results indicate that many whole-body, joint-level, reach-related, and geometric measurements, such as stature, sitting height, shoulder and elbow heights, limb lengths, and reach envelopes, were feasible across all six robots, despite differences in size and morphology. In contrast, measurements dependent on biologically specific or soft-tissue landmarks (e.g., iliac spine height, tibial height, thorax depth at nipple, bitragion arc) were infeasible for all platforms (Figure 5).

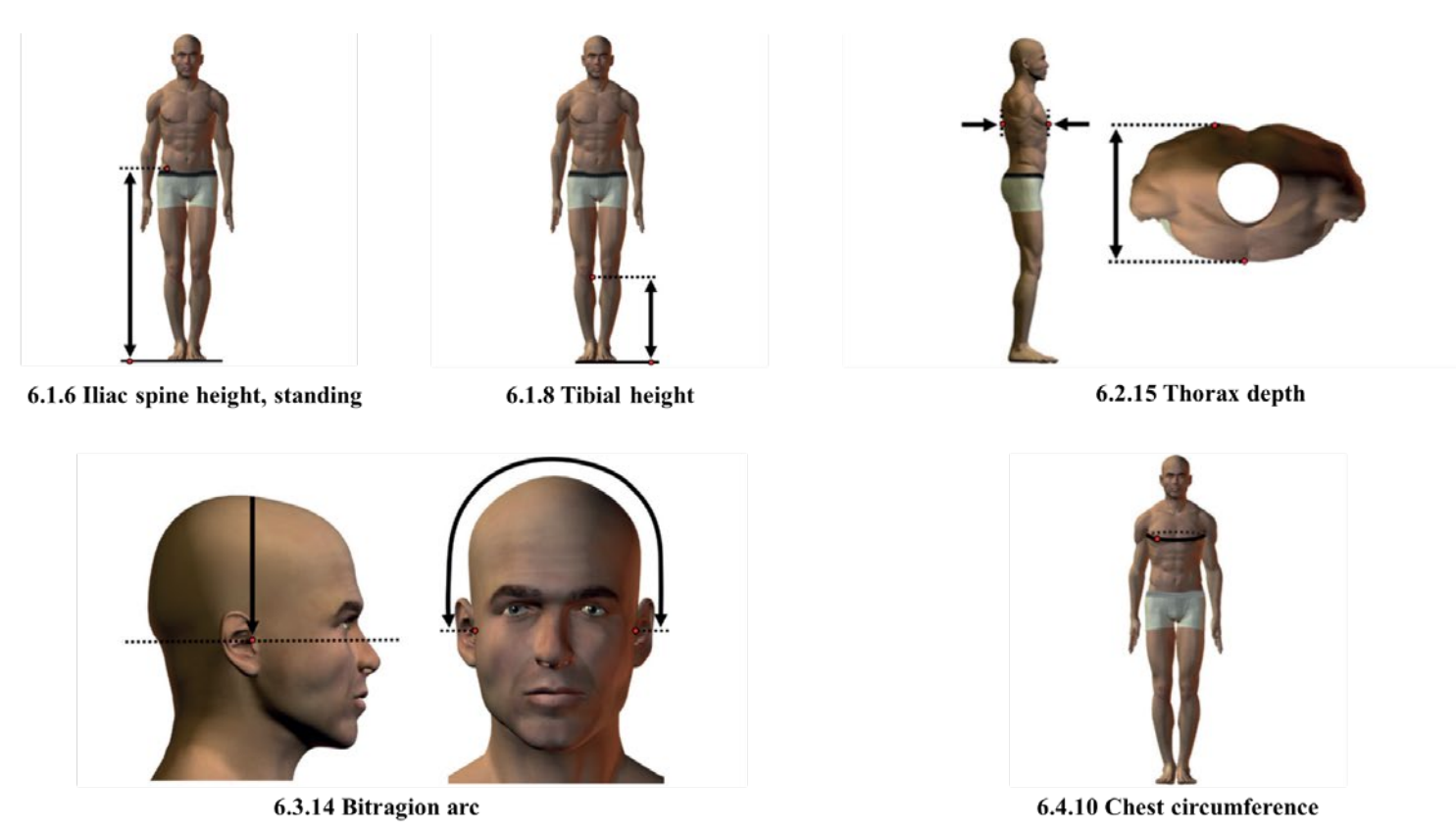

Figure 5: Measurements not measurable for any robot.

Facial-based measurements were feasible only for robots with anthropomorphic facial geometry (Figure 6). Overall, feasibility was governed by external geometry rather than human resemblance, confirming that ISO 7250 can be operationalized as a geometric reference framework but not a universally applicable anthropometric standard for humanoid robots.

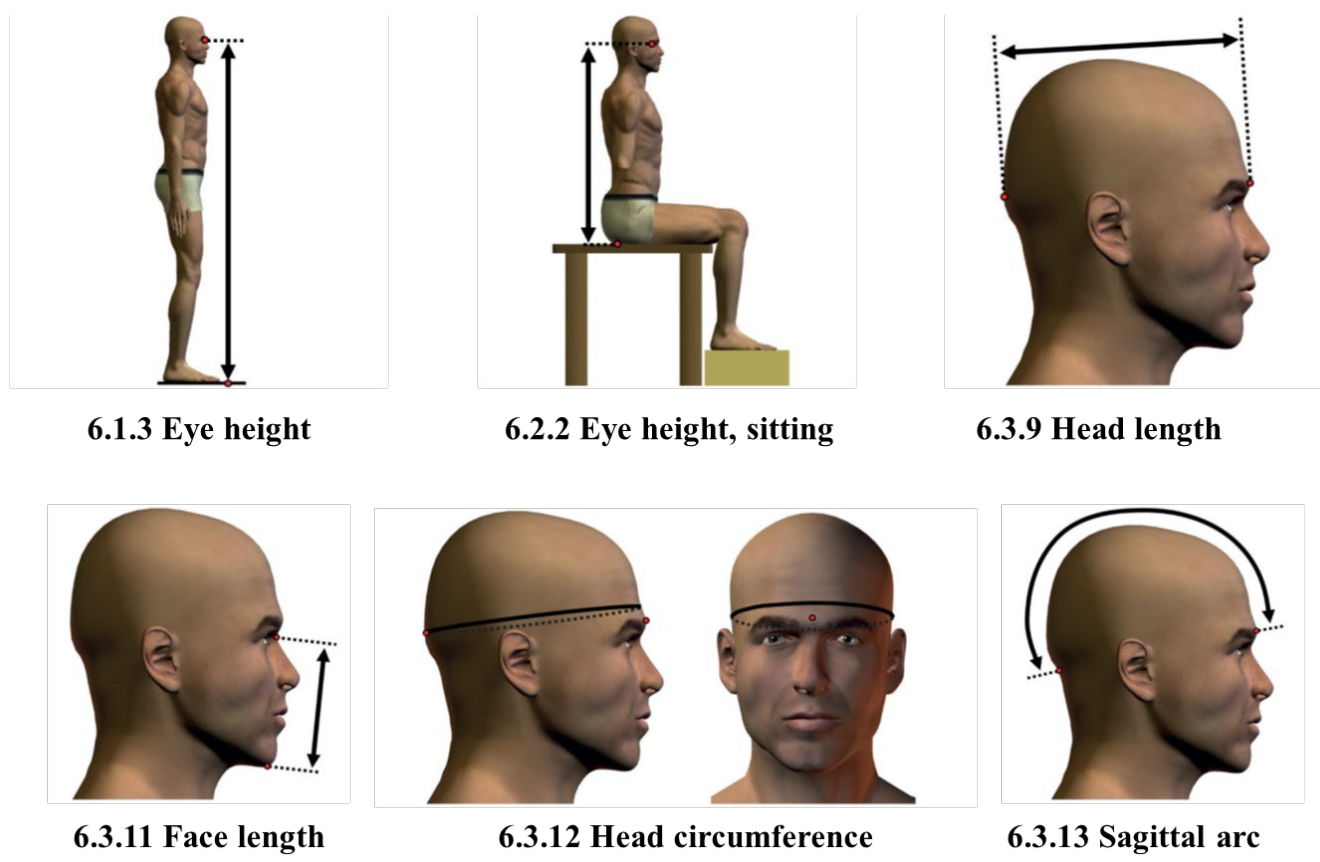

Figure 6: Measurements measurable only for Ameca.

### 3.3 Observed Deviations

Comparative analysis of the six humanoid robots revealed four recurring patterns of deviation from human anthropometric reference: additive, subtractive, exaggerative, and speculative. Taking Digit as an example (Figure 7), multiple forms of anthropometric deviation from human reference can be observed:

i. Additive deviation appears in the leg supporter, which humans apparently do not have.
ii. Subtractive deviation is evident in the absence of identifiable facial landmarks, such as nose, mouth, and ears.
iii. Exaggerative deviation is reflected in the eyes being noticeably too large and having a square shape.
iv. Speculative deviation is most apparent in the backward bending of the leg, like a bird's leg, which follows a non-human anatomical logic optimized for locomotion.

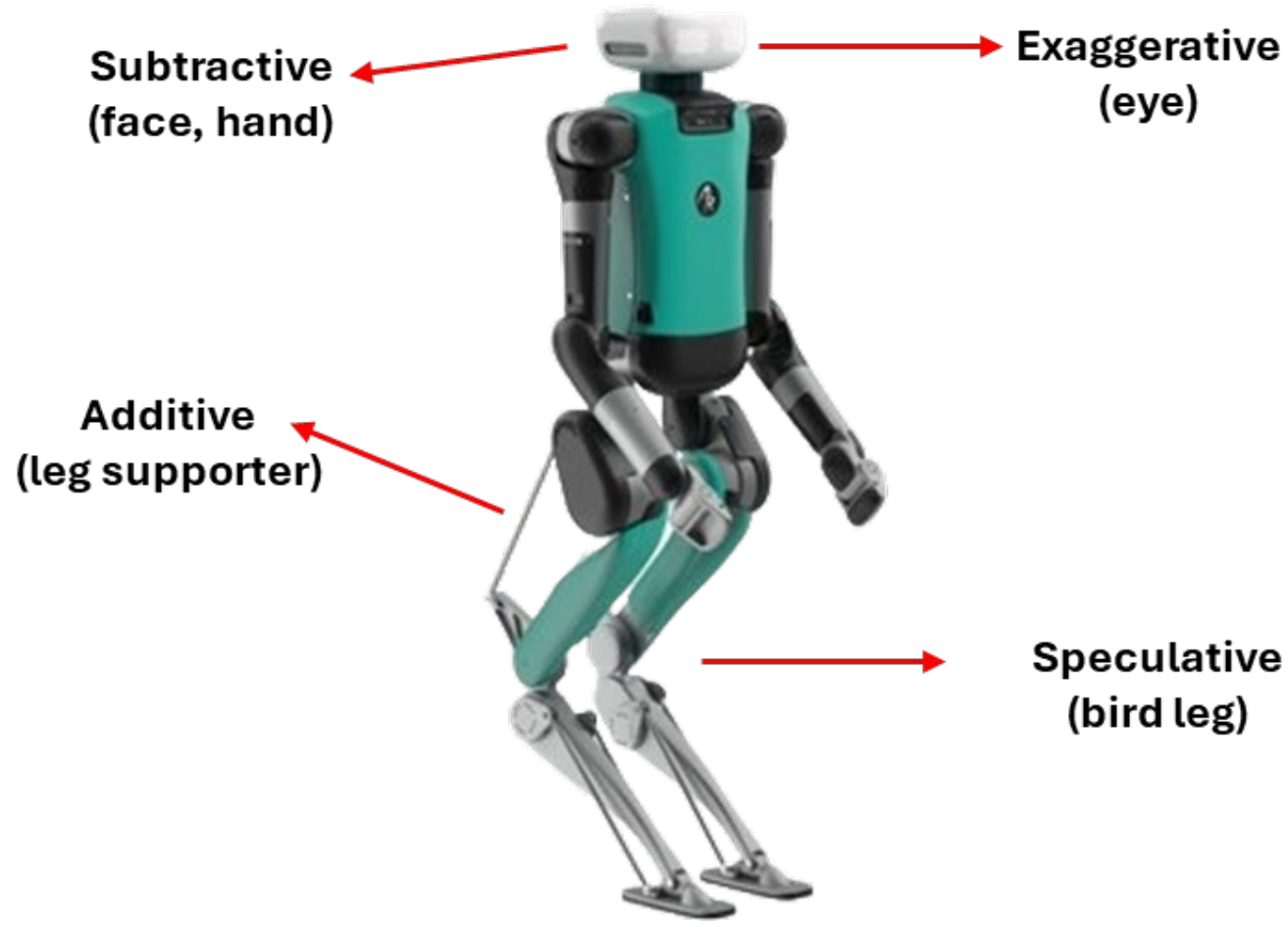

Figure 7: Observed patterns of anthropometric deviation.

These patterns often co-occur within a single platform and collectively describe how humanoid embodiments diverge from human anthropometric assumptions. Such deviations alter clearance requirements, reach alignment, motion predictability, and human risk perception during close interaction, providing a structural explanation for ergonomic risk in human–humanoid collaboration.

### 3.4 Quantitative Comparison

Quantitative comparisons were conducted across the six humanoid robots (N = 6). Biologically dependent landmarks (e.g., ASIS, thelion, tragion) were non-identifiable in 100% of cases, whereas most joint-based and geometric landmarks remained identifiable or proxy-identifiable (70–85%). Facial landmarks were identifiable in only one platform (≈17%), indicating limited anthropomorphic fidelity. Cross-platform variability was evaluated using normalized anthropometric ratios. The coefficient of variation (CV) for key proportions (e.g., arm-length-to-stature ratio) was approximately 10–15%, exceeding typical intra-human variability (3–5%). This indicates that variability among humanoid platforms may surpass natural human variation. Moreover, all robots exhibited at least two deviation types, with additive and subtractive deviations present across all platforms. Robots with higher anthropomorphic fidelity showed greater landmark coverage, while function-oriented designs showed reduced compatibility with human anthropometric assumptions.

These results highlight a structural mismatch between biological anthropometry and humanoid embodiment. Missing anatomical landmarks limit the applicability of standard ergonomic assessment methods (e.g., RULA/REBA), while deviations in body proportions and geometry may affect reach alignment, clearance, and posture during collaboration. Unlike human-centered design, which accommodates population variability, each humanoid represents a fixed anthropometric configuration, potentially increasing adaptation demands and ergonomic strain on human workers. Due to the limited sample size (N = 6), formal inferential testing was not emphasized; however, the consistent patterns across platforms indicate that anthropometric deviation is systematic and design-driven.

## 4. Conclusions

This study examined ergonomic risk in human–humanoid collaboration from an anthropometric perspective by benchmarking six contemporary humanoid robots against ISO 7250 human body measurement definitions using externally observable geometry. The results show that while many humanoid robots operate at a human scale, they exhibit systematic differences in landmark availability, body proportions, and external geometry relative to human anthropometric assumptions. The contribution of this study is testable and reproducible, as the benchmarking procedure can be applied to additional humanoid platforms or extended using motion capture and posture data in future work. Overall, ergonomic risk in human–humanoid collaboration arises not from size alone, but from how and where humanoid bodies diverge from human anthropometric logic. These findings highlight the need for anthropometry-aware evaluation approaches and greater transparency in robot dimensional disclosure to support safer system design and integration.

## Acknowledgement

The author gratefully acknowledges the financial support provided by the Bailey Faculty Fellowship for this research.